\documentclass[11pt]{article}
\usepackage[margin=1in]{geometry}
\usepackage[utf8]{inputenc}
\usepackage{physics}
\usepackage{amsmath}
\usepackage{amssymb}
\usepackage{amsfonts}
\usepackage{amsthm}
\usepackage{graphicx}

\usepackage{url}

\usepackage[superscript]{cite}

\usepackage[super,numbers,sort&compress]{natbib}
\usepackage{graphicx}% Include figure files
\usepackage{dcolumn}% Align table columns on decimal point
\usepackage{bm}% bold math
\usepackage{xr-hyper}
\makeatletter
\newcommand*{\addFileDependency}[1]{% argument=file name and extension
  \typeout{(#1)}
  \@addtofilelist{#1}
  \IfFileExists{#1}{}{\typeout{No file #1.}}
}
\makeatother

\title{\vspace{-5em}Core packing of  well-defined x-ray and NMR structures is the same}

\date{\vspace{-5em}}

\usepackage[affil-it]{authblk} 
\usepackage{etoolbox}
\usepackage{lmodern}

\makeatletter
\patchcmd{\@maketitle}{\LARGE \@title}{\fontsize{16}{19.2}\selectfont\@title}{}{}
\makeatother

\author[1,2]{Alex T. Grigas}
\affil[1]{Graduate Program in Computational Biology and Bioinformatics, Yale University, New Haven, Connecticut, 06520, USA}
\affil[2]{Integrated Graduate Program in Physical and Engineering Biology, Yale University, New Haven, Connecticut, 06520, USA}
\author[2,3]{Zhuoyi Liu}
\affil[3]{Department of Mechanical Engineering and Materials Science, Yale University, New Haven, Connecticut 06520, USA}
\author[4]{Lynne Regan}
\affil[4]{Institute of Quantitative Biology, Biochemistry and Biotechnology, Centre for Synthetic and Systems Biology, School of Biological Sciences, University of Edinburgh}
\author[1,2,3,5,6]{Corey S. O'Hern}
\affil[5]{Department of Physics, Yale University, New Haven, Connecticut 06520, USA}
\affil[6]{Department of Applied Physics, Yale University, New Haven, Connecticut 06520, USA}

\begin{document}

\maketitle

\textbf{}

\textbf{Manuscript Pages: 12}

\textbf{Total Manuscript Figures:} 6

\textbf{Total Manuscript Tables:} 0

\textbf{Supporting Information Pages: 12}

\textbf{Total Supporting Information Figures/Tables: 7/1}

\textbf{Abstract:} Numerous studies have investigated the differences and similarities between protein structures determined by solution NMR spectroscopy and those determined by x-ray crystallography. A fundamental question is whether any observed differences are due to differing methodologies, or to differences in the behavior of proteins in solution versus in the crystalline state. Here, we compare the properties of the hydrophobic cores of high-resolution protein crystal structures and those in NMR structures, determined using increasing numbers and types of restraints. Prior studies have reported that many NMR structures have denser cores compared to those of high-resolution x-ray crystal structures. Our current work investigates this result in more detail, and finds that these NMR structures tend to violate basic features of protein stereochemistry, such as small non-bonded atomic overlaps and few Ramachandran and side chain dihedral angle outliers. We find that NMR structures solved with more restraints, and which do not significantly violate stereochemistry, have hydrophobic cores that have a similar size and packing fraction as their counterparts determined by x-ray crystallography at high-resolution. These results lead us to conclude that, at least regarding the core packing properties, high-quality structures determined by NMR and x-ray crystallography are the same, and the differences reported earlier are most likely a consequence of methodology, rather than fundamental differences between the protein in the two different environments.

\textbf{Significance:} Dense packing of hydrophobic residues is key to protein structure and stability. Previously, it has been noted that structures solved by NMR spectroscopy have denser cores than x-ray crystal structures. Here, we calculate the core size and packing fraction of NMR structures with experimental restraints in the PDB. Their cores are typically smaller, but denser. However, NMR structures with accurate stereochemistry possess core packing properties that are nearly identical to high-resolution x-ray crystal structures. 

\textbf{Availability: \url{https://github.com/agrigas115/NMR_code}}

\textbf{Keywords:} hydrophobic core $|$ protein structure $|$ protein design

\newpage

\section{\label{sec:intro}Introduction}

We seek to further clarify the key physical properties of proteins that determine their structure and function. This physics-based approach is complementary to that of the remarkably successful recent machine learning methods that predict protein structure from sequence~\cite{ml:PereiraProteins2021,ml:JumperNature2021}. We focus our analyses on protein hydrophobic cores, because the core region essentially defines the folded structure and thermodynamic stability of proteins~\cite{folding:DillBiochemistry1990,folding:PaceJMB2011}. 

Over the past three decades, protein structures have been determined primarily by x-ray crystallography and solution NMR spectroscopy. Prior studies have reported differences between structures determined by NMR and those determined by x-ray crystallography~\cite{nmrvxtal:GarbuzynskiyProt2005,nmrvxtal:AndrecProteins2007,nmrvxtal:SchneiderProt2009,nmrvxtal:SikicBiochemJour2010,nmrvxtal:MaoJACS2014,nmrvxtal:EverettProtSci2016,nmrvxtal:KoehlerProt2018,subgroup:MeiProteins2020}. A fundamental question is whether the observed differences are due to differing methodologies, or to differences in the behavior of proteins in solution versus in the crystalline state. 

For structures determined by x-ray crystallography, it is universally accepted that a good structure has a low R-factor and that high-resolution structures represent a more accurate picture of the structure than low-resolution structures. In previous work, we analyzed a non-redundant set of high-resolution x-ray crystal structures (with resolution $< 1.8$~\AA) and found that the cores (residues with zero solvent accessible surface area) of these proteins represent about $8\%$ of the total number of residues in the protein. In addition, the packing fraction, $\phi$, in protein cores is $\sim 0.55 \pm 0.01$~\cite{subgroup:GainesPRE2016,subgroup:GrigasProSci2020}. In earlier work, using a subset of NMR structures with a large number of distance restraints per residue, we concluded that the cores of NMR-determined protein structures had a higher packing fraction than those of high-resolution protein structures determined by x-ray crystallography~\cite{subgroup:MeiProteins2020}. Denser core packing in NMR structures has also been previously reported in other studies~\cite{nmrvxtal:RatnaparkhiBiochem1998,nmrvxtal:GarbuzynskiyProt2005}. However, for structures determined by NMR, there is no equivalent universal definition of resolution or metric of quality~\cite{nmrqual:MontelioneStruc2013}. Here, we filter the NMR dataset by a wide array of structural metrics and track how the physical properties of the protein core vary.

Currently, there are approximately $12$,$000$ protein structures solved via solution-NMR spectroscopy available in the Protein Data Bank (PDB). In this article, we focus on those structures that have restraints accessible via the PDB, which is roughly half. We first analyzed the core packing of all individual structures in these NMR bundles. We found that on average there is a lower fraction of residues in the core, $f_{c}$ , than in high-resolution x-ray crystal structures. However, the packing fraction, $\phi$, of the cores of protein structures determined by NMR is higher than that of the cores determined by x-ray crystallography. Moreover, these results also hold when comparing the structures of the same protein, determined by both NMR and x-ray crystallography.

To investigate the possible origins of the differences seen in NMR core packing compared to high-resolution x-ray crystal structures, we filtered the dataset by a variety of structural metrics. We find that NMR bundles solved with over $10$ distance restraints per residue have a fraction of residues in the core that is similar to that of high-resolution x-ray crystal structures, but the average packing fraction of those cores is higher than that observed in the cores of protein structures determined by x-ray crystallography. We investigated this phenomenon further by selecting NMR structures that satisfy common stereochemical metrics, such as low Clashscores and few Ramachandran and sidechain dihedral angle outliers. Applying these filters results in only $72$ bundles of the original $6$,$449$. The $f_{c}$ and packing fraction of the remaining $72$ bundles is the same as that observed in high-resolution protein x-ray crystal structures. These results suggest that protein structures are fundamentally the same in solution as in crystals, and that previously noted differences in the core are a consequence of the structure-determination methodology.

\section{Results}
\label{results}

First, we assembled datasets of x-ray crystal and NMR protein structures from the PDB. For the x-ray crystal structures, we used a dataset of $5$,$261$ high-resolution x-ray crystal structures culled from the PDB using PISCES~\cite{PISCES:WangBioinformatics2003,PISCES:WangNucleicAcids2005} with resolution $< 1.8$~\AA, sequence identity cutoff of $< 20\%$ and R-factor cutoff of $< 0.25$. For the NMR structures, we retrieved all NMR bundles from the PDB that have available restraint data (i.e. a subset of distance restraints derived from the nuclear Overhauser effect (NOE), dihedral angle restraints from J-couplings, and bond-vector orientational restraints from residual dipolar couplings (RDCs)) using \textsc{nmr2gmx}~\cite{nmr2gmx:SinelnikovaJBNMR2021}, resulting in $6$,$449$ NMR bundles with a total of $127$,$959$ individual structures. We refer to these $127$,$959$ NMR structures as the entire NMR dataset. We subsequently filter this dataset according to the number of NOE distance restraints per residue and by several stereochemical validation metrics. Our goal is to understand how the core properties depend on these parameters.

We define core residues as those whose relative solvent accessible surface area, rSASA, is $\leq 10^{-3}$. We calculate the packing fraction, $\phi$ , of core residues, which is defined as the ratio of the volume of the residue to the volume of the Voronoi polyhedron that encloses it~\cite{subgroup:TreadoPRE2019}. We have shown in previous studies of high-resolution x-ray crystal structures that $\phi$ for core residues is $\sim 0.55 \pm 0.01$. (We also found that $\phi$ increases with decreasing rSASA~\cite{subgroup:GainesProteins2018}, until it reaches a plateau at $\phi \sim 0.55$ for rSASA $< 10^{-3}$.) (See Methods for more details.) 

\begin{figure}[ht]
\centering
\includegraphics[width=0.85\textwidth]{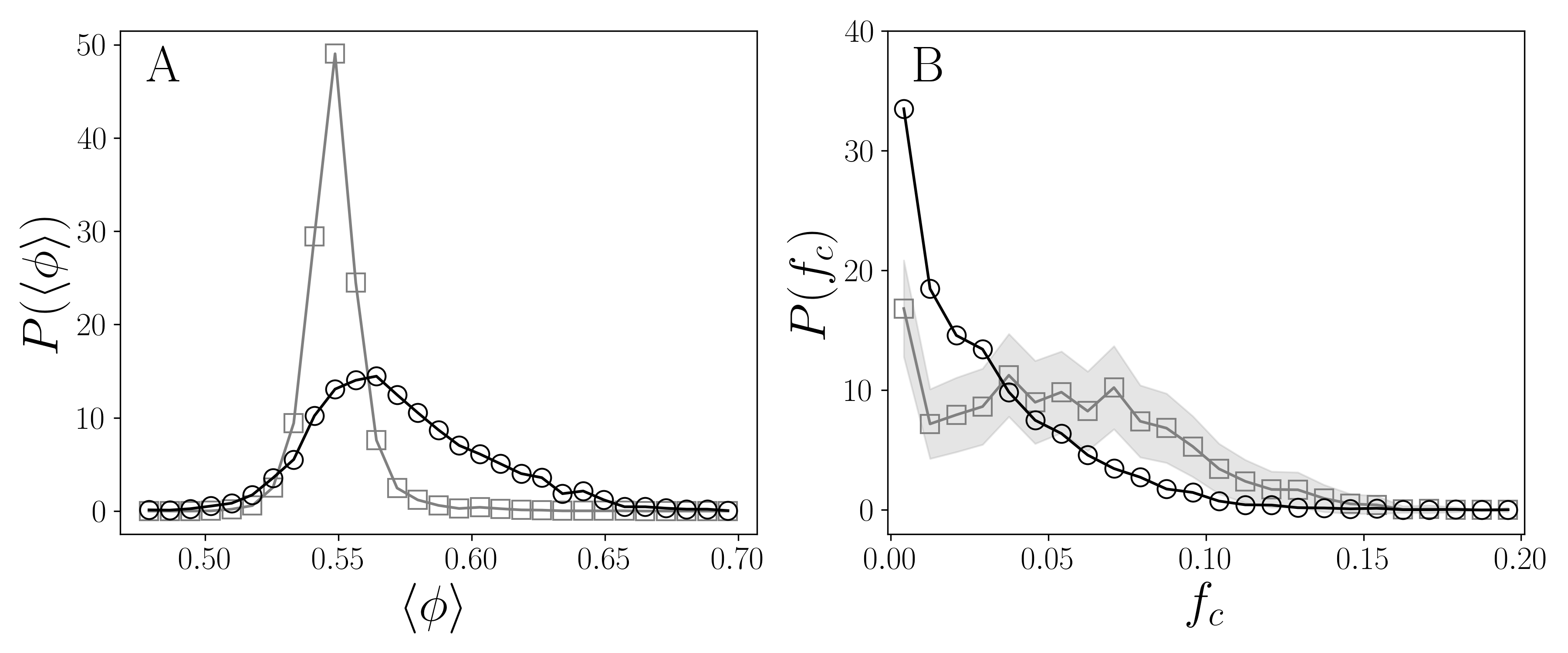}
\caption{(A) The distribution of the packing fraction of core residues $P(\langle \phi \rangle)$ for high-resolution x-ray crystal structures (grey squares) and NMR structures (black circles). For the x-ray crystal structures, $\langle \phi \rangle$ is averaged over all core residues for each protein and for the NMR structures, $\langle \phi \rangle$ is averaged over all core residues for all structures in the bundle for each protein. (B) The distribution $P(f_{c})$ for the resampled x-ray crystal structures (grey squares with shading based on the standard deviation) averaged over $10^3$ resampling trials and for NMR structures (black circles), where the protein size is limited to $50 < N < 300$.}
\label{fig:nmr_packing}
\end{figure}

In Fig.~\ref{fig:nmr_packing} (A), we compare the distributions of the average packing fraction $\langle \phi \rangle$ for core residues in the high-resolution x-ray crystal structure dataset  and the entire NMR dataset. For the x-ray crystal structures, we average over the core residues in each protein structure, resulting in a single value of $\langle \phi \rangle$ per protein. For the NMR structures, $\langle \phi \rangle$ is calculated by averaging over all core residues across all structures in each bundle,  resulting in a single $\langle \phi \rangle$ per protein. We find that for the entire NMR dataset the average value of $\langle \phi \rangle$ is  $0.57$, (two standard deviations above the value of $0.55$ for x-ray crystal structures), and it is clear in Fig.~\ref{fig:nmr_packing} (A) that the distribution of $\langle \phi \rangle$ is very broad, and distinctly skewed towards more densely packed structures.

The average packing fraction in the core, $\langle \phi \rangle$,  is essentially independent of the total number of residues in the protein, $N$. In contrast, the fraction of the protein that is core, $f_{c}$, increases with increasing $N$. (See Supporting Information.) Because $f_{c}$  depends on $N$, it is important to compare datasets with the same length distribution. The proteins in the NMR data set tend to be smaller than those in the dataset of x-ray crystal structures, therefore to calculate $P(f_{c})$, we resampled the x-ray crystal dataset so that it possesses a $P(N)$ that is similar to that for the NMR dataset for $N < 300$. The resampling was obtained by averaging over $10^{3}$ independent trials. $P(f_{c})$ for the NMR dataset and the resampled x-ray crystal structure dataset with $N < 300$ is plotted in Fig. 1 (B). It is clear that $P(f_{c})$ has a strong peak for $f_{c} \sim 0$ (i.e. no core residues) for the NMR dataset. In contrast, $P(f_{c})$ shows that most x-ray crystal structures of these lengths have a core of $f_{c} \sim 0.05$ on average. In summary, when we analyze all structures in the PDB determined by NMR, for which restraint data is available, they on average possess smaller, more densely packed cores compared to high-resolution x-ray crystal structures. Interestingly, smaller, but more densely packed cores have also been found in many computationally generated low-scoring structures with incorrect backbone placement submitted to the Critical Assessment of protein Structure Prediction (CASP) competitions~\cite{subgroup:GrigasProSci2020}.

\begin{figure}[h]
\centering
\includegraphics[width=0.55\textwidth]{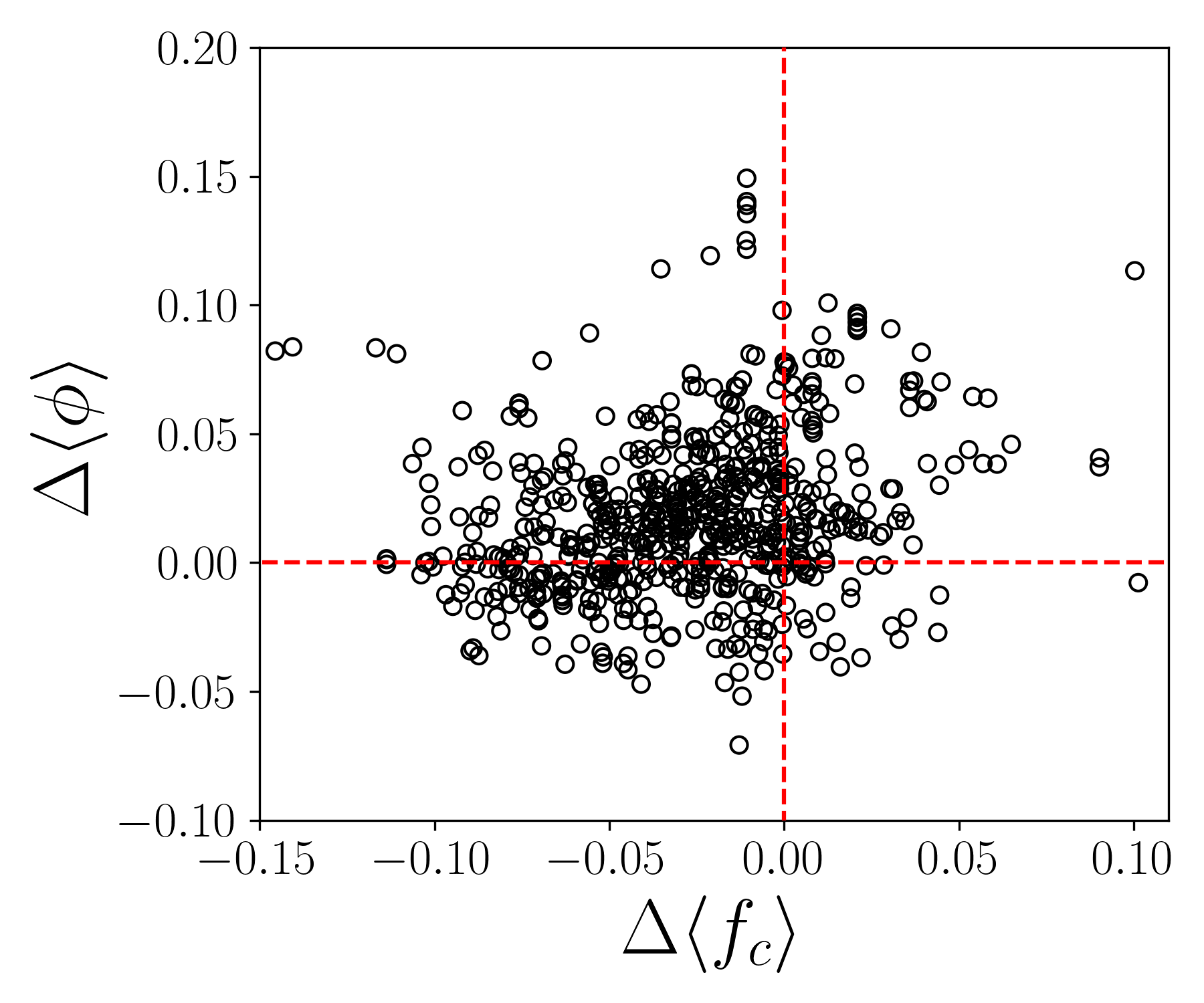}
\caption{Scatter plot of the differences in the packing fraction $\Delta \langle \phi \rangle$ and fraction of core resides $\Delta f_c$ between the NMR and x-ray crystal structures in the paired dataset.}
\label{fig:nmr_xtal_pair_packing}
\end{figure}

To investigate the discrepancies in the core packing properties in greater detail, we compared protein structures that had been determined by both x-ray crystallography and NMR. We assembled a dataset of $702$ pairs with both a high-resolution x-ray crystal structure ($< 2.0$~\AA) and an NMR structure with available restraints, where the pairs have $> 90\%$ sequence similarity. (See Methods.) In Fig.~\ref{fig:nmr_xtal_pair_packing}, we show a scatter plot of the differences in the average core packing fraction and fraction of core residues between the NMR and x-ray crystal structures, where $\Delta \langle \phi \rangle = \langle \phi \rangle_{\rm NMR} - \langle \phi \rangle_{\rm x-ray}$ and $\Delta \langle f_{c} \rangle= \langle f_{c} \rangle_{\rm{NMR}} - \langle f_{c} \rangle_{\rm x-ray}$. Consistent with Fig.~\ref{fig:nmr_packing}, the majority of NMR structures have smaller, more densely packed cores compared to their x-ray crystal structure counterparts, i.e. most data points occur in the upper left corner. However, there is also data in the three other quadrants. 

\begin{figure}[h]
\centering
\includegraphics[width=\textwidth]{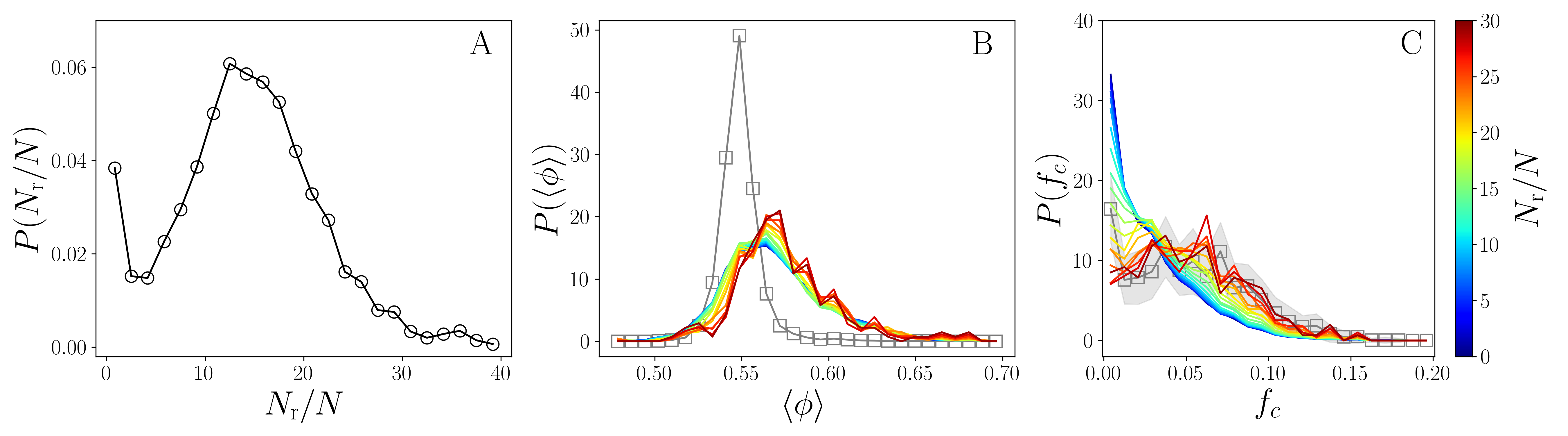}
\caption{(A) The distribution $P(N_{\rm{r}}/N)$ of the number of distance restraints per residue $N_{\rm{r}}/N$ for the NMR dataset. (B) The distribution $P(\langle \phi \rangle)$ of the average core packing fraction and (C) distribution $P(f_c)$ of the fraction of core residues for NMR structures with $N_{\rm{r}}/N$ greater than a given value (indicated by the color from $0$ in violet to $30$ in dark red). In (B) and (C), we also show $P(\langle \phi \rangle)$ and $P(f_c)$ for the resampled high-resolution x-ray crystal dataset (grey squares with shading that indicates one standard deviation).}
\label{fig:packing_vs_frac_res}
\end{figure}

These results raise the interesting question of whether the differences in the size and packing fraction of the core represent true differences between protein structures when they are in crystalline states versus in solution? Or are the differences a consequence of the methods used to determine the atomic structure? Since NMR bundles are determined by a set of restraints (i.e. distance restraints derived from NOEs, dihedral angle restraints from J-couplings, and bond-vector orientational restraints from RDCs), it is reasonable to assume that the more restraints used, the higher the quality of the NMR structure. Note that while requiring a large number of restraints is necessary, previous studies have shown that it is not sufficient for identifying high-quality NMR structures~\cite{nmrqual:SpronkJBioNMR2003,nmrqual:ZagrovicProteins2006,nmrqual:RosatoCurrOpinStrucBio2013}. Nevertheless, it is of interest to track how the core properties vary with the number of NOE measurements per residue. In a typical NMR structure there are many more NOE distance restraints than any other type of restraint, so we chose to analyse their effect on core properties.

We show  the distribution $P(N_{\rm{r}}/N)$ of the number of NOE distance restraints per residue in the NMR dataset in Fig.~\ref{fig:packing_vs_frac_res} (A). There are many NMR structures with $N_{\rm{r}}/N \sim 1$, but the vast majority have $N_{\rm{r}}/N \sim 10$-$25$, and a few have $N_{\rm{r}}/N \gtrsim 30$. In Fig.~\ref{fig:packing_vs_frac_res} (B), we plot the distribution of the average packing fraction of core residues, $P(\langle \phi \rangle)$, in NMR structures with different values of $N_{\rm{r}}/N$. We find that $P(\langle \phi \rangle)$ does not depend strongly on $N_{\rm{r}}/N$ and differs from $P(\langle \phi \rangle)$ for high-resolution x-ray crystal structures for all values of $N_{\rm{r}}/N$. In contrast, the distribution of the average fraction of core residues, $P(f_c)$, in NMR structures becomes increasingly similar to that of the high-resolution x-ray crystal structure dataset for $N_{\rm{r}}/N \gtrsim 20$, as shown in Fig.~\ref{fig:packing_vs_frac_res} (C). There is no correlation between $N$ and $N_{\rm{r}}/N$ and therefore we used the same resampling of the x-ray crystal structure database in Fig.~\ref{fig:nmr_packing}. The number of NOE distance restraints per residue has been reported to be only an approximate proxy of the quality of the protein structure, since the NOE distance restraints can be redundant~\cite{noeinfo:NabuursJACS2003}. However, when we perform the same analyses for the more informative inter-residue distance restraints involving side chain atoms, the results are the same. (See the Supporting Information.)

\begin{figure}[h]
\centering
\includegraphics[width=0.95\textwidth]{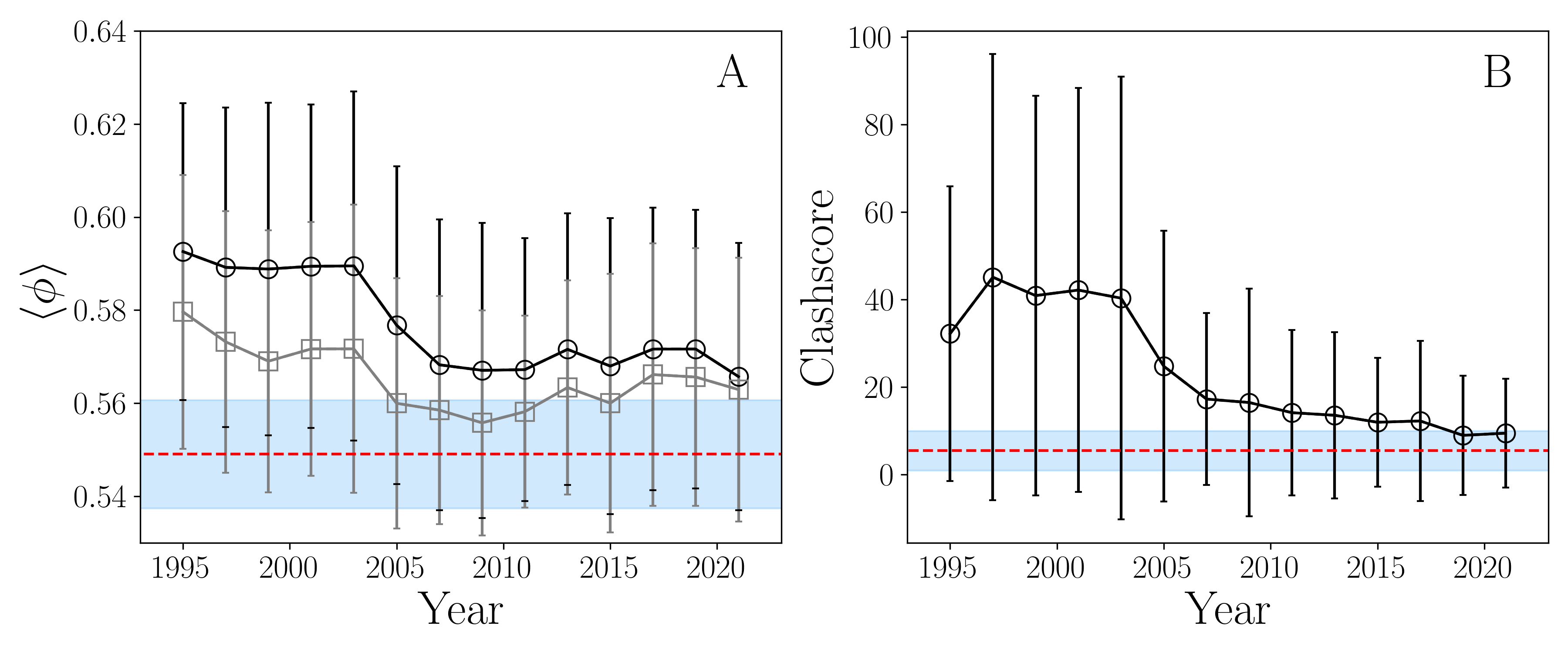}
\caption{(A) The average packing fraction $\langle \phi \rangle$ of core residues and (B) Clashscore in NMR structures versus the year that they were deposited in the PDB for the entire NMR dataset (black circles) and those with a Clashscore that is lower than the average value for high-resolution x-ray crystal structures plus one standard deviation (grey squares). The average values and standard deviations of $\langle \phi \rangle$ and Clashscore for high-resolution x-ray crystal structures are indicated by red dashed lines and blue shading, respectively.}
\label{fig:packing_over_time}
\end{figure}

We were interested in investigating if the packing fraction of core residues in NMR structures changed over time as the methodologies for NMR structure determination have improved~\cite{nmrqual:VuisterJBioNMR2014,nmrqual:BuchnerStructure2015,molprobity:WilliamsProSci2018}. In Fig.~\ref{fig:packing_over_time} (A), we show the average packing fraction of core residues $\langle \phi \rangle$ versus the year that the NMR structures were deposited in the PDB. We find that  $\langle \phi \rangle$ was largest before $2004$ ($\langle \phi \rangle \sim 0.59$), decreased to $\langle \phi \rangle \sim 0.57$ after 2004, and has remained nearly constant over the past decade. One possible cause of the elevated values of $\langle \phi \rangle$ in the NMR structures, relative to the value of $0.55$ observed in high-resolution x-ray crystal structures, is overlap of non-bonded atoms, which can increase the apparent packing fraction. It is therefore of interest to note that the Clashscore, defined by the MolProbity software~\cite{molprobity:WilliamsProSci2018} as the number of non-bonded atomic overlaps that are greater than $0.4$~\AA~ per $1$,$000$ atoms, has decreased over time for NMR structures, with a rapid decrease around $2004$ (the year MolProbity was first published) and then a slow, but steady decrease towards the average value for high-resolution x-ray crystal structures over the last decade.  (See  Fig.~\ref{fig:packing_over_time} (B).) The wwPDB NMR Validation Task Force has emphasized that validation metrics, such as the Clashscore and Ramachandran and side chain dihedral angle outliers, used to determine the quality of x-ray crystal structures are appropriate for assessing NMR structures~\cite{nmrqual:MontelioneStruc2013}. In particular, these metrics should be similar for all high-quality protein structures, i.e. structures that are obtained from x-ray crystallography and solution NMR spectroscopy~\cite{nmrvalidation:SpronkProgNMR2004,nmrqual:VuisterJBioNMR2014}. When we remove NMR structures that have a Clashscore larger than the average value plus one standard deviation of the Clashscore for high-resolution x-ray crystal structures (i.e. Clashscore $> 10.5$), we find that $\langle \phi \rangle$ decreases significantly, but $\langle \phi \rangle$ for these NMR structures with minimal or no overlap of non-bonded atoms, remains above the value for high-resolution x-ray crystal structures. 

These results indicate that while removing NMR structures that contain unphysical atomic overlaps accounts for some of the overpacking found in NMR structures, $\langle \phi \rangle$ for NMR structures is still above the value for x-ray crystal structures. Thus, there is another reason, in addition to atomic overlaps, that is responsible for the elevated packing fraction observed in some NMR structures. Although many of the metrics are inter-related, we considered several additional quality metrics, and analysed how the properties of the cores of the protein structures depend on them. To this end, we explicitly consider the fraction of Ramachandran and  sidechain dihedral angle outliers, bond length and bond angle outliers, fraction of backbone chemical shifts assigned, and strength of the magnet used. If we filter the NMR dataset using each of these metrics one at a time, $\langle \phi \rangle$ for NMR structures decreases toward the value for high-resolution x-ray crystal structures and the $P(f_{c})$ distribution for NMR structures more closely resembles that of the x-ray crystal structure distribution. However, whether or not the structure was solved with RDCs makes little difference for the core packing properties. (See Supporting Information.)

\begin{figure}[h]
\centering
\includegraphics[width=0.95\textwidth]{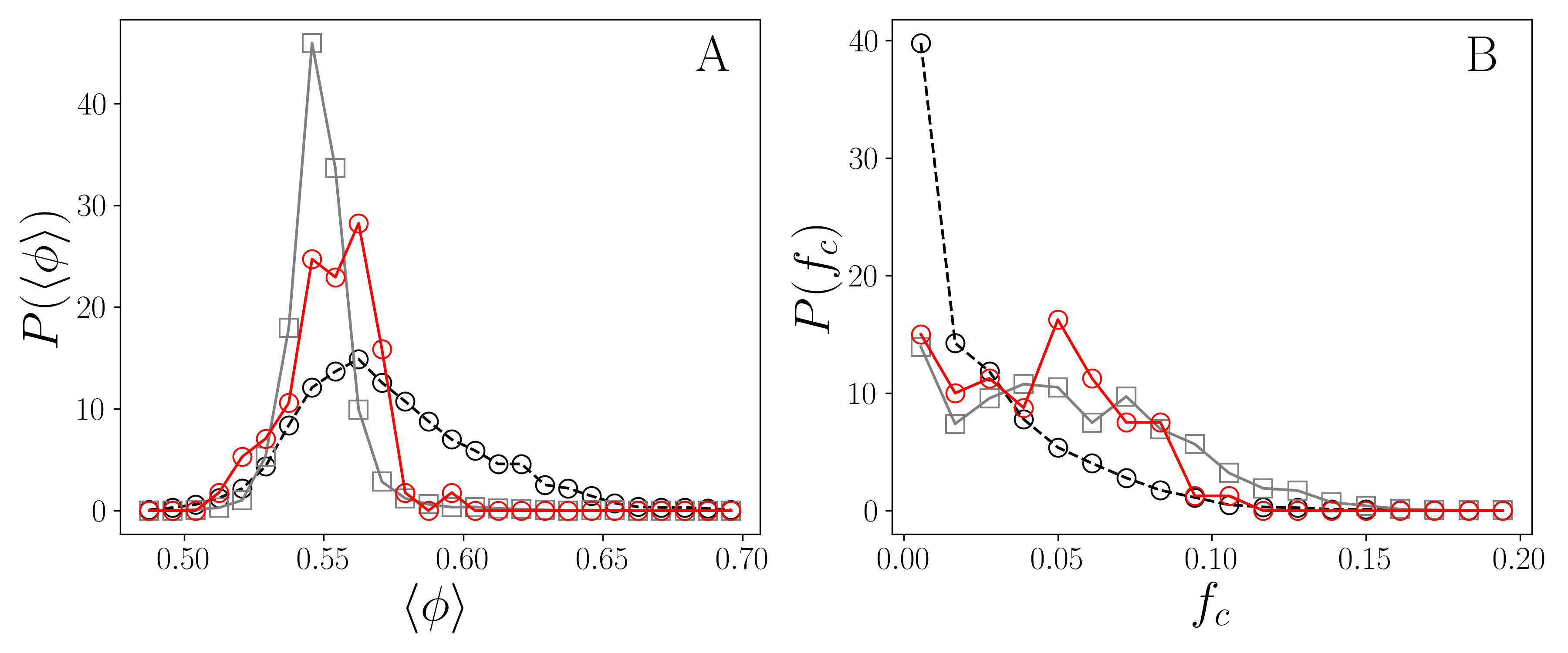}
\caption{(A) The distribution $P(\langle \phi \rangle)$ of the average packing fraction for core residues and (B) the distribution $P(f_{c})$ of the fraction of core residues for high-resolution x-ray crystal structures (grey squares), the filtered NMR structures (red circles), and the entire NMR dataset (black circles dashed).}
\label{fig:nmr_packing_filtered}
\end{figure}

For determining the validation metric cutoffs, we sought a compromise between strict cutoffs near the high-resolution x-ray crystal structure values and loose cutoffs that admit more data. When we filter the NMR dataset so that the NMR structures 1) possess values for the Clashscore, Ramachandran and sidechain dihedral angle outliers that are within one standard deviation of the average for the high-resolution x-ray crystal structures (i.e. Clashscore $< 10.5$, Ramachandran dihedral angle outliers $< 0.5\%$ and side chain dihedral angle outliers $< 3\%$), 2) have at least $10$ NOE distance restraints per residue, 3) have $60\%$ or more backbone chemical shifts assigned, and 4) were solved on a $650$ MHz magnet or stronger, $\langle \phi \rangle$ and $P(f_{c})$ for the NMR structures are nearly identical to those for the high-resolution x-ray crystal structures as shown in Fig.~\ref{fig:nmr_packing_filtered}. 

These filters are quite stringent, as the filtered dataset contains only $72$ NMR structures of the $6$,$449$ from the entire NMR dataset. The average packing fraction of core residues in this filtered NMR dataset is $\langle \phi \rangle = 0.55 \pm 0.02$ compared to the value of $\langle \phi \rangle = 0.55 \pm 0.01$ for the high-resolution x-ray crystal structure dataset. The distribution of the fraction of core residues is also similar for the filtered NMR structures and high-resolution x-ray crystal structures. (See Figure~\ref{fig:nmr_packing_filtered}.) Similar results can be achieved when only the stereochemical criteria are considered. (See Supporting Information.) None of the filtered NMR bundles is included the x-ray/NMR paired dataset.

\begin{figure}[h]
\centering
\includegraphics[width=\textwidth]{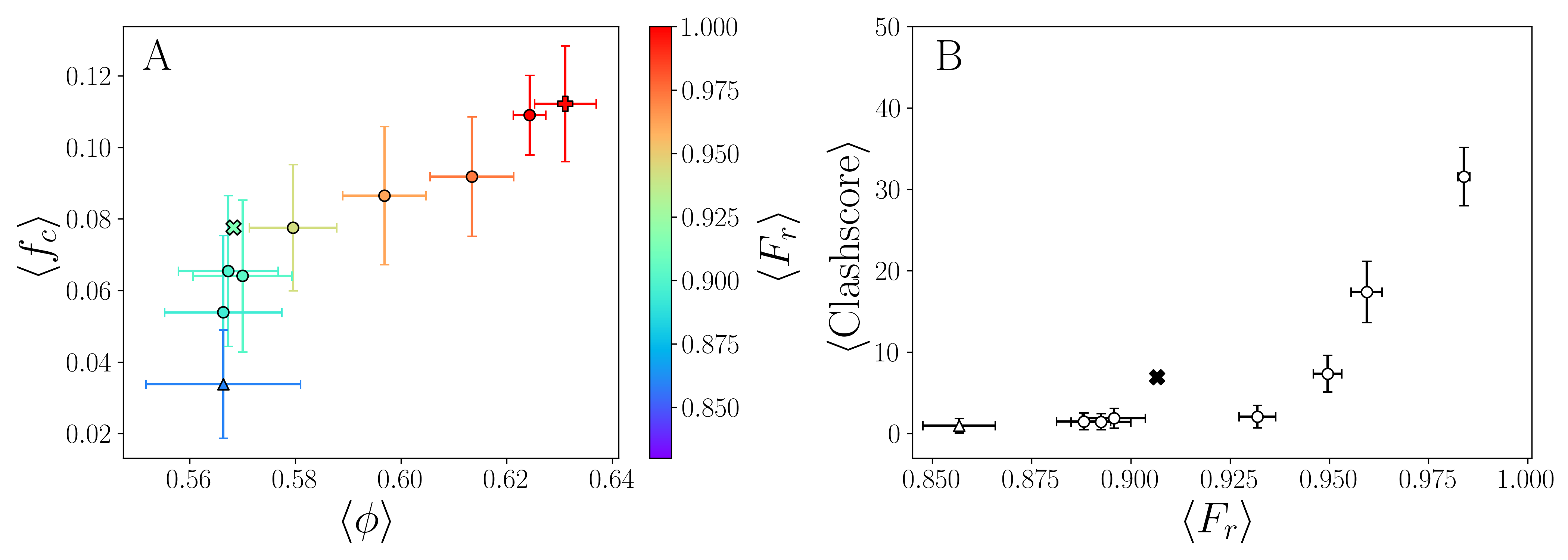}
\caption{(A) Average packing fraction of core residues $\langle \phi \rangle$ and average fraction of core residues $\langle f_{c} \rangle$ for the x-ray crystal structure with PDBID: 2CRW (x), the NMR bundle with PDBID: 2CZN (+), and unrestrained (upward triangle) and restrained (circles) MD simulations colored by the average fraction of NOE distance restraints $\langle F_r \rangle$ that are satisfied. (B) The average Clashscore plotted versus the average fraction of NMR distance restraints satisfied $\langle F_{r} \rangle$ for the systems in (A). The data for the NMR bundle with a high Clashscore of $124$ is not shown.}
\label{fig:rMD}
\end{figure}

To investigate the origin of the deviation in the packing properties between NMR and x-ray crystal structures, we used GROMACS to carry out all-atom MD simulations of a globular protein, the chitin binding domain of chitinase from Pyrococcus furiosus, which has an x-ray crystal structure with a resolution of $1.7$~\AA~(PDBID: 2CWR) and an NMR bundle with a large number of distance restraints (PDBID: 2CZN). There are $38$ individual models in the NMR bundle, solved with an average of $25$ distance restraints per residue. Nevertheless, the bundle has a high Clashscore of $124$. We used MD simulations to investigate if we can remove the atomic overlaps in the NMR bundle and match the NMR distance restraints to the same extent as the deposited bundle. When retrieving NOE distance restraints from the PDB, the restraints are reported as a lower bound, a first upper bound, and a second upper bound on the pairwise atomic distances. Here, we define a distance restraint as being satisfied if the distance is less than the first upper bound. In the case of ambiguous NOE restraints, where the NOE restraint is attributed to more than a single pair of atoms (e.g. methyl hydrogens), we count the restraint as satisfied if at least one of the set of pairs has a distance less than the first upper bound. To assess the resulting structures, we then calculated $\langle \phi \rangle$ and $\langle f_{c} \rangle$ over the MD simulation trajectory.

In Fig.~\ref{fig:rMD} (A), we plot the average fraction of core residues $\langle f_{c} \rangle$ versus the average packing fraction of core residues $\langle \phi \rangle$ from the MD simulations and for the x-ray crystal and NMR structures. For this particular protein, the NMR bundle has both a higher packing fraction and a larger fraction of core residues than the x-ray crystal structure, but it has a large Clashscore and many Ramachandran and sidechain dihedral angle outliers. The color scale in Fig.~\ref{fig:rMD} (A) indicates the average fraction $F_{r}$ of the distance restraints that are satisfied (using the first upper bound). 

We first ran unrestrained MD simulations, starting from a structure in the NMR bundle. This data point is plotted as a triangle in Fig.~\ref{fig:rMD} (A). In the unrestrained MD simulations, $85\%$ of the distance restraints are satisfied compared to $\sim 93\%$ in the x-ray crystal structure (x) and $\sim 100\%$ in the NMR bundle (+). Note that in the unrestrained MD simulations $\langle \phi \rangle$ decreases from $\langle \phi \rangle \sim 0.63$ for the initial NMR structure to the value for the x-ray crystal structure, $\langle \phi \rangle \sim 0.57$. However, the fraction of core residues, $\langle f_{c} \rangle \sim 0.03$, is much smaller than that for the initial NMR bundle with $\langle f_{c} \rangle \sim 0.11$ and the x-ray crystal structure with $f_{c} \sim 0.08$. The unrestrained MD simulation was run for $500$ ns, during which time the packing properties rapidly equilibrate, and therefore the restrained MD simulations described below were run for shorter times. (See Supporting Information.) 

We also ran MD simulations with restraints to recapitulate the NOE distances. GROMACS enforces NMR distance restraints using the following potential energy:
\begin{equation} 
\label{eq:dist_rest}
V_{dr}(r_ {ij})  = \begin{cases}
\frac{k}{2}(r_{ij}-r_{0})^2, & r_{ij}<r_0,\\
0, & r_{0} \leq r_{ij} < r_1,\\
\frac{k}{2}(r_{ij} - r_1)^2, & r_1 \leq r_{ij} < r_2,\\
\frac{k}{2}(r_2-r_1)(2r_{ij}-r_2 - r_1), & r_2 \leq r_{ij},\\
\end{cases}
\end{equation}
where $r_{ij}$ is the separation between NOE atom pairs $i$ and $j$, $r_0$, $r_1$, and $r_2$ are the lower bound, first upper bound, and second upper bound distances for the NOE atom pairs. Additionally, the average separation $\langle r_{ij}\rangle$ is weighted by a factor of $r^{-3}_{ij}$ multiplied by an exponentially decaying function of time with decay constant $\tau$. (See Methods.) This restraint potential is similar to the potentials used in NMR structure determination software~\cite{xplor:SchwietersJMagRes2003,xplor:SchwietersProgNucMag2006,cyana:Guntert2004}. Each restrained simulation was initialized using a single structure from the NMR bundle and run for $20$ ns following energy minimization, and the NOE pair separations and packing properties were obtained every $0.1$ ns and averaged. We considered several spring constants $k = 10^3$, $10^4$, and $10^5$ kJ/mol/nm$^2$ and a decay time $\tau=0.5$~ns, which increases the average fraction $\langle F_r \rangle$ of distance restraints that are satisfied from $0.86$ to $0.90$. For $k > 10^5$ kJ/mol/nm$^2$, $\langle F_{r} \rangle$ saturates and cannot reach $\langle F_{r} \rangle \sim 1$, which is the value for the NMR bundle deposited in the PDB. (See Supporting Information.) Restrained MD simulations using the flat-bottomed potential in Eq.~\ref{eq:dist_rest} yield cores with similar values of $\langle f_c \rangle$ and $\langle \phi \rangle$ as the x-ray crystal structure. To obtain $\langle F_{r} \rangle \sim 1$, we implemented a new flat-bottomed distance restraint potential, where $r_1$ is set half way between the original lower bound and original first upper bound, using spring constants $k = 10^3$, $10^4$, and $10^5$ kJ/mol/nm$^2$ and a decay time $\tau=0.5$~ns and a final highly restrained simulation with a spring constant $k = 10^5$ kJ/mol/nm$^2$ and $\tau=0$~ns. (See Methods.) With these modifications to the restraint potential and $\tau=0$~ns, we are able to obtain structures that have $\langle F_r \rangle \sim 1$ and values of $\langle f_c \rangle$ and $\langle \phi \rangle$ that match those for the NMR bundle. However, in Fig.~\ref{fig:rMD} (B), we show that as $\langle F_r\rangle$ increases toward $1$, the Clashscore rises steeply,  reaching values that are indicative of a poor quality protein structure. In addition, as shown in the Supporting Information, the same behavior is found for the fraction of Ramachandran and side chain dihedral angle outliers. The final increase in $\langle F_{r} \rangle$ toward $1$ is caused by the increase in the number of satisfied inter-residue distance restraints involving side chain atoms. (See the Supporting Information.) By enforcing the distance restraints such that $\langle F_r \rangle \sim 1$ in the MD simulations, we obtain the same core packing properties as the deposited NMR bundle, but these structures violate metrics of protein stereochemistry. Thus, matching all of the deposited NOE restraints to the same extent as the deposited NMR bundle is likely the cause of the overpacking observed in this particular NMR bundle.

\section{Discussion}
\label{discussion}

An important question in protein science is to what extent are the structures of proteins in the crystalline state different from those of proteins in solution. Here, we have shown that most NMR structures deposited in the PDB possess smaller cores, and that these cores are more densely packed than those observed in high-resolution x-ray crystal structures. More densely packed cores in NMR structures have also been reported previously for smaller datasets~\cite{subgroup:MeiProteins2020,nmrvxtal:GarbuzynskiyProt2005,nmrvxtal:RatnaparkhiBiochem1998}. However, we find that by only considering NMR bundles that have a large number of restraints and that pass protein stereochemistry validation metrics (i.e. those that have small Clashscores and few Ramachandran and sidechain dihedral angle outliers), both the core size and packing fraction of these NMR structures match the values found for high-resolution x-ray crystal structures. The validation metrics are quite stringent: Only $72$ NMR bundles of the original $6$,$449$ bundles in the dataset pass the validation metrics. We interpret these results to show that the packing properties of the cores of proteins in the crystalline state (determined by high-resolution x-ray crystallography) and the cores of proteins in solution (determined by state-of-the-art NMR spectroscopy) are the same. However, since none of the filtered NMR structures have high-resolution x-ray pairs, it remains possible that particular proteins may  have non-zero $\Delta \langle \phi \rangle$ or $\Delta f_{c}$.

A recent study reported that many NMR structures have surface loops that are often under-restrained and too floppy~\cite{ansurr:FowlerStruc2021}. This observation is consistent with our finding that the cores of NMR structures solved with few distance restraints are typically too small. In particular, inappropriately positioned surface residues will affect the solvent accessibility of the interior residues.

Over-packing occurs simultaneously with poor protein stereochemistry (large non-bonded atomic clashes and many dihedral angle outliers) in NMR structures in the PDB. However, this result does not rule out the possibility that denser core packing than that found in high-resolution x-ray crystal structures could occur in principle without violating protein stereochemistry. For example, packings of amino acid-shaped particles can become denser with the addition of thermal fluctuations, in contrast to packings that are rapidly compressed and energy-minimized~\cite{subgroup:MeiProteins2020}. Therefore, we carried out MD simulations with NOE distance restraints to attempt to remove non-bonded atomic overlaps and dihedral angle outliers and then measured the resulting core packing properties. Our MD simulations on a particular globular protein, the chitin binding domain of chitinase from Pyrococcus furiosus, showed that forcing the NOE distance restraints to be satisfied to the same extent as the deposited NMR bundle leads to incorrect core packing features, non-bonded atomic clashes, and dihedral angle outliers. 

The NOE restraint data clearly contains important structural information, as the NMR and the x-ray crystal structures have similar overall folds (with an average C$_\alpha$ RMSD $=1.6$~\AA) and loosely enforcing all of the NOE restraints in the MD simulations leads to core packing properties that are similar to the high-resolution x-ray crystal structure. However, enforcing the NOE distance restraints to the same extent of the deposited NMR bundle appears to be inconsistent with protein steroechemistry. The fact that simultaneously satisfying all of the NOE distance restraints leads to violations of protein stereochemistry (for this particular protein) is likely related to the fact that the conversion of NOE measurements to distance bounds is approximate and NOEs report on time-averaged atomic separations, which can be challenging to implement in protein structure determination~\cite{nmrqual:ZagrovicProteins2006,noes:VogeliProgNucMag2014,noes:ChalmersJMagRes2016}. We emphasize that our MD simulations were performed on a single protein whose structure was determined by both x-ray crystallography and NMR (from quadrant 2 of Fig.~\ref{fig:nmr_xtal_pair_packing}).  In future studies, it will be important to study protein structure pairs that represent all four quadrants from Fig.~\ref{fig:nmr_xtal_pair_packing} and to determine what restrained MD protocol would be needed to improve the modeling of the protein core. Additionally, our unrestrained MD simulations show that the core packing fraction is recapitulated using the Amber99SB-ILDN forcefield and TIP3P water model, but the core size is not. Therefore, we will explore whether the same result is observed with different force fields and solvation models, which would indicate how well the hydrophobic effect is currently captured in MD simulations of proteins. 

\section{Materials and Methods}
\label{sec:MatMethods}

\subsection*{Datasets}

We collected two datasets of protein structures from the Protein Data Bank (PDB). We first compiled a dataset of $5$,$261$ high-resolution x-ray crystal structures from the PDB using PISCES~\cite{PISCES:WangBioinformatics2003,PISCES:WangNucleicAcids2005} with resolution $< 1.8$~\AA, a sequence identity cutoff of $< 20\%$, an R-factor cutoff of $< 0.25$, and lengths greater than $50$ residues. We also assembled a large dataset of NMR structures from the PDB. We identified all NMR bundles that have available restraint data in the NMR-STAR format using \textsc{nmr2gmx}~\cite{nmr2gmx:SinelnikovaJBNMR2021}, resulting in $6$,$499$ NMR bundles, each with at least $10$ structures in the bundle. In this dataset, $5$,$830$ bundles have NOE distance restraints, $3$,$994$ have dihedral angle restraints, and $507$ have RDC restraints. The NMR dataset was not filtered for similarity in amino acid sequence, and thus it has multiple bundles for some proteins. \textsc{PyPDB} was used to retrieve all of the validation metrics for each protein from the PDB~\cite{pypdb:GilpinBioinfo2015}. In addition, \textsc{PyPDB} was used to identify a set of NMR and x-ray crystal structure pairs. We selected pairs for which the x-ray crystal structure sequence matched $> 90\%$ via MMSeqs2~\cite{mmseqs2:SteineggerNatBiotech2017}, the lengths differ by $< 10 \%$, the x-ray crystal structure resolution is $\leq 2.0$~\AA, and the NMR bundle has restraint data. This dataset has $702$ pairs, made up of $514$ x-ray crystal structures and $525$ NMR structures, as some NMR structures have multiple x-ray crystal structure matches and vice versa.

\subsection*{rSASA}

To identify core residues, we measured each
residue's solvent accessible surface area (SASA). To calculate SASA,
we use the \textsc{Naccess} software
package~\cite{naccess:Hubbard1993}, which implements an algorithm
originally proposed by Lee and Richards~\cite{rsasa:LeeJMB1971}. To normalize the
SASA, we take the ratio of the SASA within the context of
the protein ($\text{SASA}_{\text{context}}$) and the SASA of the same
residue (X) extracted from the protein structure as a dipeptide
(Gly-X-Gly) with the same backbone and side-chain dihedral angles:
\begin{equation}
\text{rSASA} = \frac{\text{SASA}_{\text{context}}}{\text{SASA}_{\text{dipeptide}}}.
\end{equation}
Core residues are classified as those that have ${\rm rSASA} \leq
10^{-3}$. The choice of this threshold has been discussed in several previous studies~\cite{subgroup:GainesProteins2018,subgroup:TreadoPRE2019}. 

\subsection*{Packing Fraction}

A characteristic measure of the packing in an atomic system is
the packing fraction. The packing fraction of residue $\mu$ is defined by
\begin{equation}
\phi_{\mu} = \frac{\nu_{\mu}}{V_{\mu}},
\end{equation}
where $\nu_{\mu}$ is the non-overlapping volume of residue $\mu$ and $V_{\mu}$
is the volume of the Voronoi cell surrounding residue $\mu$. The volume of the residue is determined by the atom sizes and their locations. The particular atom sizes that we used have been previously selected from the range of literature values to reproduce the backbone and side chain dihedral angle distributions found in high-resolution x-ray crystal structures~\cite{subgroup:ZhouBPJ2012,subgroup:ZhouProteins2014,subgroup:CaballeroProtSci2014}. The Voronoi cell
represents the local free space around the residue. We calculate the non-overlapping residue volume with a grid-based volume estimation. To
calculate the Voronoi polyhedra for a protein structure, we use surface Voronoi tessellation, which defines a Voronoi cell as the
region of space in a given system that is closer to the bounding
surface of the residue than to the bounding surface of any other
residue in the system. We calculate the surface Voronoi tessellations
using the \textsc{Pomelo} software package~\cite{pomelo:WeisEPJ2017}. This
software approximates the bounding surfaces of each residue by
triangulating points on the residue surfaces. We find that using $\sim 400$
points per atom, or $\sim 6400$ surface points per residue, gives an
accurate representation of the Voronoi cells and the results
do not change if more surface points are included.

\subsection*{Molecular Dynamics Simulations}
The unrestrained and restrained all-atom molecular dynamics (MD) simulations of the globular protein, the chitin binding domain of chitinase from Pyrococcus furiosus, were performed using the GROMACS software package~\cite{gromacs:AbrahamSoftwareX2015}. The MD simulations used a cubic box with edge lengths $63$~\AA, which ensures that the protein is $> 10$~\AA~away from the box edges, and the box is filled with water molecules modeled using TIP3P~\cite{tip3p:JorgensenJACS1981} at neutral pH and $0.15$M NaCl. Periodic boundary conditions are applied in the $x$-, $y$-, and $z$-directions. Short-range van der Waals and screened Coulomb interactions were truncated at $1.2$ nm, while longer-ranged electrostatics were tabulated using the Particle Mesh Ewald summation method. We performed two short energy minimization runs to relax the protein first and then the water molecules and protein together using the steepest decent method. We then simulated the system for $20$ ns in the NPT ensemble at temperature $T=300$K and pressure $P=1$ bar using the weakly coupled Berendsen thermostat and barostat and calculated the $\langle \phi \rangle$ and $f_{c}$ every $0.1$~ns. The time constant of the Berendsen thermostat was set to $2$ ps and the isothermal compressibility for the Berendsen barostat was set to $4.5\times10^{-5} \rm bar^{-1}$. (See the Supporting Information for more details about the equilibration procedure.) The equations of motion for the atomic coordinates and velocities were integrated using a leapfrog algorithm with a $1$~fs time step. For both the unrestrained and restrained MD simulations, we used the AMBER99SB-ILDN force field~\cite{a99sb-ildn:BestJPhysChemB2009,a99sb-ildn:Lindorff-LarsenProteins2010}. For the restrained MD simulations, we implemented the flat-bottomed spring potential $V_{dr}(r_{ij})$ given in Eq.~\ref{eq:dist_rest} with spring constants $k = 10^3$, $10^4$, and $10^5$ kJ/mol/nm$^2$ to enforce the NOE distance restraints between atom pairs $i$ and $j$. We calculate the time average NOE pairwise atomic separations as:
\begin{equation}
\langle r_{ij} \rangle = \left \langle \left( r^{-3}_{ij}(t-\Delta t) e^{(-\Delta t/\tau)} + r^{-3}(t) \left(1-e^{(-\Delta t/\tau)}\right) \right)^{-1/3} \right\rangle_t,
\end{equation}
where $\Delta t$ is the time step and $\tau$ is the decay time. $\tau=0.5$~ns for all restraint simulations, except for the most highly restrained simulation. This time-averaging method has been suggested to more accurately represent NOE distance restraints and allows for larger fluctuations in the atomic separations if NOE distance restraints are incompatible.
To further increase the fraction of NOE restraints that are satisfied during the MD simulations to $\langle F_r \rangle \sim 1$, we also implemented a narrower flat-bottomed spring potential, where the first upper bound $r_{1}$ is set to the separation between the NOE atom pairs of the initial NMR structure, and $\tau =0$ when calculating $\langle r_{ij} \rangle$.

\newpage

%\bibliography{nmr_packing}

\end{document}